\documentclass[12pt]{article}
\usepackage[dvips]{graphicx}

\begin{document}

\begin{flushright}
{\small
SLAC--PUB--8883\\
June 2001\\}
\end{flushright}

\vfill

\begin{center}
{{\bf\LARGE
Two-Photon Processes at Intermediate\\[1.2ex] Energies}\footnote{Work
supported by the Department of Energy
under contract number DE-AC03-76SF00515.}}

\bigskip
Stanley J. Brodsky\\
{\sl Stanford Linear Accelerator Center \\
Stanford University, Stanford, California 94309\\
sjbth@slac.stanford.edu}\\
\medskip
\end{center}

\vfill

\begin{center}
{\bf\large
Abstract }
\end{center}

Exclusive hadron production processes in photon-photon collisions
provide important tests of QCD at the amplitude level,
particularly as measures of hadron distribution amplitudes and
skewed parton distributions. The determination of the shape and
normalization of the distribution amplitudes has become
particularly important in view of their importance in the analysis
of exclusive semi-leptonic and two-body hadronic $B$-decays.
Interesting two-photon physics, including doubly-tagged $\gamma^*
\gamma^*$  reactions, will be accessible at low energy,  high
luminosity $e^+ e^-$ colliders, including measurements of channels
important in the light-by-light contribution to the muon $g$--2
and the study of the transition between threshold production
controlled by low-energy effective chiral theories and the domain
where leading-twist perturbative QCD becomes applicable.  The
threshold regime of hadron production in photon--photon and $e^+
e^-$ annihilation, where hadrons are formed at small relative
velocity, is particularly interesting as a test of low energy
theorems, soliton models, and new types of resonance production.
Such studies will be particularly valuable in double-tagged
reactions where polarization correlations, as well as the photon
virtuality dependence, can be studied.

\vfill

\begin{center}
{\it Invited talk presented at the\\  $e^+ e^-$ Physics}
{\it at Intermediate Energies Workshop}\\
{\it Stanford Linear Accelerator Center, Stanford, California}\\
{\it April 30--May 2, 2001}
\end{center}

\vfill

\newpage

\renewcommand{\baselinestretch}{1.1}
\normalsize

\section{Introduction}
Two-photon annihilation $\gamma^*(q_1) \gamma^*(q_2) \to $ hadrons
for real and virtual photons provide some of the most detailed and
incisive tests of QCD.  Among the processes of special interest
are:
\begin{enumerate}
\item the total two-photon annihilation hadronic cross section
$\sigma(s,q^2_1,q^2_2),$ which is related to the light-by-light
hadronic contribution to the muon anomalous moment;

\item the formation of $C = +$ hadronic resonances, which can
reveal
 exotic states such as $q
\bar q g$ hybrids and discriminate gluonium
formation \cite{Pennington:2000ai,Acciarri:2001ex};

\item single-hadron processes such as $\gamma^* \gamma^* \to
\pi^0$, which test the transition from the anomaly-dominated pion
decay constant to the short-distance structure of currents
dictated by the operator-product expansion and perturbative QCD
factorization theorems;

\item hadron pair production processes such as $\gamma^* \gamma \to
\pi^+ \pi^-, K^+ K^-, p \bar p$, which at fixed invariant pair
mass measures the $s \to t$ crossing of the virtual Compton
amplitude  \cite{Brodsky:1981rp,BrodskyLepage}.   When one photon
is highly virtual, these exclusive hadron production channels are
dual to the photon structure function $F^\gamma_2(x,Q^2)$ in the
endpoint $x \to 1$ region at fixed invariant pair mass.  The
leading twist-amplitude for $\gamma^* \gamma \to \pi^+ \pi^-$ is
sensitive to the $1/x - 1/(1-x)$ moment of the $q \bar q$
distribution amplitude $\Phi_{\pi^+ \pi^-}(x,Q^2)$ of the two-pion
system \cite{Muller:1994fv,Diehl:2000uv}, the timelike extension
of skewed parton distributions.  In addition one can measure the
pion charge asymmetry in $e^+ e^- \to \pi^+ \pi^- e^+ e^-$ arising
from the interference of the $\gamma \gamma \to \pi^+ \pi^-$
Compton amplitude with the timelike pion form factor
\cite{Brodsky:1971ud}.  At the unphysical point $s =q^2_1= q^2_2 =
0$, the amplitude is fixed by the low energy theorem to the hadron
charge squared.  As reviewed by Karliner in these proceedings
\cite{Karliner}, the ratio of the measured $\gamma \gamma \to
\Lambda \bar \Lambda$ and $\gamma \gamma \to p \bar p$ cross
sections is anomalous at threshold, a fact which may be associated
with the soliton structure of baryons in QCD
\cite{Sommermann:1992yh};

\item At large momentum transfer, the angular distribution of
hadron pairs produced by photon-photon annihilation are among the
best determinants of the shape of the meson and baryon
distribution amplitudes $\phi_M(x,Q)$,  and $\phi_B(x_i,Q)$ which
control almost all exclusive processes involving a hard scale $Q$.
The determination of the shape and normalization of the
distribution amplitudes, which are gauge-invariant and
process-independent measures of the valence wavefunctions of the
hadrons, has become particularly important in view of their
importance in the analysis of exclusive semi-leptonic and two-body
hadronic $B$-decays
\cite{BHS,Sz,BABR,Beneke:1999br,Keum:2000ph,Keum:2000wi}. There
has also been considerable progress both in calculating hadron
wavefunctions from first principles in QCD and in measuring them
using diffractive di-jet dissociation.
\end{enumerate}

Much of this important two-photon physics is accessible at low energy,
high luminosity $e^+ e^-$ colliders such as the proposed PEP-N project,
particularly for measurements of channels important in the
light-by-light contribution to the muon
$g$--2 and the exploration of the transition between threshold
amplitudes which are controlled by low-energy effective theories such as
the chiral Hamiltonian through the transition to the domain where
leading-twist perturbative QCD becomes applicable.  There have been
almost no measurements of double-tagged events needed to unravel the
separate $q^2_1$ and $q^2_2$ dependence of photon-photon annihilation.
Hadron pair production from two-photon annihilation plays a
crucial role in unraveling the perturbative and non-perturbative
structure of QCD, first by testing the validity and empirical
applicability of leading-twist factorization theorems, second by
verifying the structure of the underlying perturbative QCD subprocesses,
and third, through measurements of angular distributions and ratios which
are sensitive to the shape of the distribution amplitudes.  In effect,
photon-photon collisions provide a microscope for testing fundamental
scaling laws of PQCD and for measuring distribution amplitudes.  It would
also be interesting to measure the novel relativistic atomic coalescence
processes,  single and double muonium formation: $e^+ e^- \to
[\mu^+ e^-]\mu^+ e^-$ and $e^+ e^- \to
[\mu^+ e^-] [\mu^+ e^-]$.

\section{The Photon-to-Pion Transition Form Factor and the Pion
Distribution Amplitude}

The simplest and perhaps most elegant illustration of an exclusive
reaction in QCD is the evaluation of the photon-to-pion transition
form factor $F_{\gamma \to \pi}(Q^2)$ which is measurable in
single-tagged two-photon $ee \to ee \pi^0$ reactions. The form
factor is defined via the invariant amplitude $\Gamma^\mu = -ie^2
F_{\pi \gamma}(Q^2) \varepsilon^{\mu \nu \rho \sigma} p^\pi_\nu
\varepsilon_\rho q_\sigma$. As in inclusive reactions, one must
specify a factorization scheme which divides the integration
regions of the loop integrals into hard and soft momenta, compared
to the resolution scale $\widetilde Q$. At leading twist, the
transition form factor then factorizes as a convolution of the
$\gamma^* \gamma \to q \bar q$ amplitude (where the quarks are
collinear with the final state pion) with the valence light-cone
wavefunction of the pion \cite{BrodskyLepage}:
\begin{equation}
F_{\gamma M}(Q^2)= {4 \over \sqrt 3}\int^1_0 dx \phi_M(x,\widetilde Q)
T^H_{\gamma \to M}(x,Q^2) .
\label{transitionformfactor}
\end{equation}
The hard scattering amplitude for $\gamma\gamma^*\to q \bar q$ is
$ T^H_{\gamma M}(x,Q^2) = { [(1-x) Q^2]^{-1}}\left(1 + {\cal
O}(\alpha_s)\right)$. For the asymptotic distribution amplitude
$\phi^\mathrm{asympt}_\pi (x) = \sqrt 3 f_\pi x(1-x)$ one predicts
\cite{Brodsky:1998dh}
\begin{displaymath}
Q^2 F_{\gamma \pi}(Q^2)= 2 f_\pi \left(1
- {5\over 3} {\alpha_V(Q^*)\over \pi}\right)
\end{displaymath}
where $Q^*= e^{-3/2} Q$ is the estimated BLM scale for the pion
form factor in the $V$ scheme.

\begin{figure}[htbp]
\centering
\includegraphics[width=.69\columnwidth]{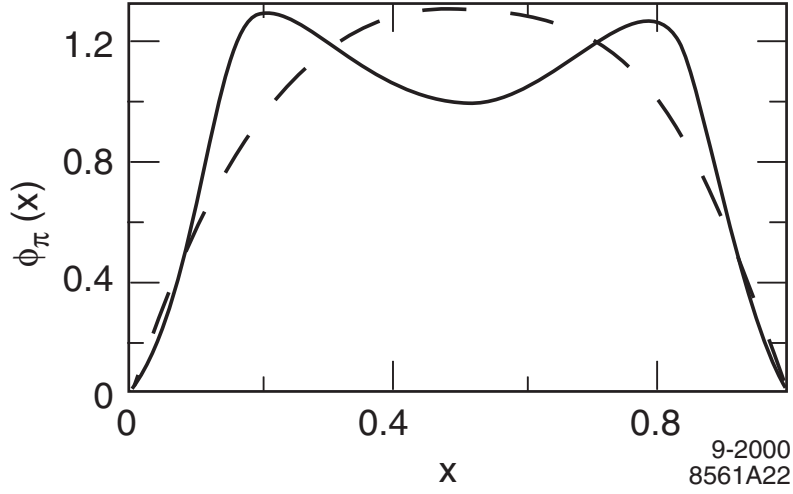}\\
(a)\\[15pt]
\includegraphics[width=.69\columnwidth]{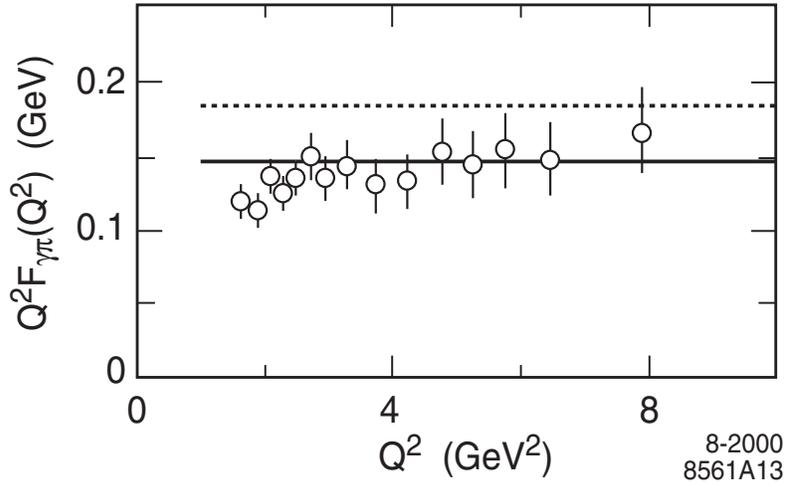}\\
(b)
\caption{
(a) Transverse lattice results for the pion distribution
amplitude at $Q^2
\sim 10 {\rm GeV}^2$.  The solid curve is the theoretical
prediction from the combined DLCQ/transverse lattice
method \cite{Dalley:2000dh}; the chain line is the experimental result
obtained from dijet
diffractive dissociation \cite{Ashery:1999nq,Aitala:2001hb}.
Both are normalized to the same area for comparison.  (b) Scaling of the
transition photon to pion transition form factor $Q^2F_{\gamma
\pi^0}(Q^2)$.  The dotted and solid theoretical curves are the
perturbative QCD prediction at leading and next-to-leading order,
respectively, assuming the asymptotic pion distribution The data are from
the CLEO collaboration \cite{Gronberg:1998fj}.}
\label{Fig:DalleyCleo}
\end{figure}

The PQCD predictions have been tested in measurements of $e \gamma
\to e \pi^0$ by the CLEO collaboration \cite{Gronberg:1998fj} (see
Figure \ref{Fig:DalleyCleo} (b)). The flat scaling of the $Q^2
F_{\gamma \pi}(Q^2)$ data from $Q^2 = 2$ to $Q^2 = 8$ GeV$^2$
provides an important confirmation of the applicability of leading
twist QCD to this process.  The magnitude of $Q^2 F_{\gamma
\pi}(Q^2)$ is remarkably consistent with the predicted form,
assuming the asymptotic distribution amplitude and including the
LO QCD radiative correction with $\alpha_V(e^{-3/2} Q)/\pi \simeq
0.12$.  One could allow for some broadening of the distribution
amplitude with a corresponding increase in the value of $\alpha_V$
at small scales. Radyushkin \cite{Radyushkin}, Ong \cite{Ong} and
Kroll \cite{Kroll} have also noted that the scaling and
normalization of the photon to pion transition form factor tends
to favor the asymptotic form for the pion distribution amplitude
and rules out broader distributions such as the two-humped form
suggested by QCD sum rules \cite{CZ}.  More comprehensive
analyses, which include consideration of next-to-leading
order corrections and some higher-twist contributions dictated by
vector meson spectra and QCD sum rules have been given by
A.~Khodjamirian \cite{Khodjamirian:1999tk},
A.~Schmedding and O.~Yakovlev \cite{Schmedding:2000ap},  and by
A.~P.~Bakulev {\em et al.} \cite{Bakulev:2001ud}.

When both photons are virtual, the
denominator of $T_H$ for the $\gamma \gamma^* \to \pi^0$ reaction becomes
$(1-x) Q^2_1 + x Q^2_2$ \cite{BrodskyLepage,Ong}, and the amplitude
becomes nearly insensitive to the shape of the distribution amplitude
once it is normalized to the pion decay constant.  Thus the ratio of
singly virtual to doubly virtual pion production is particularly
sensitive to the shape of
$\phi_\pi(x,Q^2)$ since higher order corrections and normalization errors
tend to cancel in the ratio.

\section{Non-Perturbative Calculations of the Pion Distribution Amplitude}

The distribution amplitude $\phi(x,\widetilde Q)$ can be computed from
the integral over transverse momenta of the renormalized hadron valence
wavefunction in the light-cone gauge at fixed light-cone time
\cite{BrodskyLepage}:
\begin{equation}
\phi(x,\widetilde Q) = \int d^2\vec{k_\perp}\thinspace
\theta \left({\widetilde Q}^2 - {\vec{k_\perp}^2\over x(1-x)}\right)
\psi^{(\widetilde Q)}(x,\vec{k_\perp}),
\label{quarkdistamp}
\end{equation}
where a global cutoff in invariant mass is identified with the
resolution $\widetilde Q$.  The distribution amplitude $\phi(x,
\widetilde Q)$ is boost and gauge invariant and evolves in $\ln
\widetilde Q$ through an evolution equation \cite{BrodskyLepage}.
Since it is formed from the same product of operators as the
non-singlet structure function, the anomalous dimensions
controlling $\phi(x,Q)$ dependence in the ultraviolet $\log Q$
scale are the same as those which appear in the DGLAP evolution of
structure functions \cite{Brodsky:1980ny}. The decay $\pi \to \mu
\nu$ normalizes the wave function at the origin: $\int^1_0 dx
\phi(x,Q) = {f_\pi/ (2 \sqrt 3)}$. One can also compute the
distribution amplitude from the gauge invariant Bethe-Salpeter
wavefunction at equal light-cone time.  This also allows contact
with both QCD sum rules \cite{Shifman:1979by} and lattice gauge
theory; for example, moments of the pion distribution amplitudes
have been computed in lattice gauge theory
\cite{Martinelli:1987si,Daniel:1991ah,DelDebbio:2000mq}. Conformal
symmetry can be used as a template to organize the renormalization
scales and evolution of QCD predictions
\cite{Brodsky:1980ny,Brodsky:2000cr}. For example,  Braun and
collaborators have shown how one can use conformal symmetry to
classify the eigensolutions of the baryon distribution amplitude
\cite{Braun:1999te}.

Dalley \cite{Dalley:2000dh} and Burkardt and Seal
\cite{Burkardt:2001mf} have calculated the pion distribution
amplitude from QCD using a combination of the discretized
light-cone quantization \cite{dlcq} method for the $x^-$ and $x^+$
light-cone coordinates with the transverse lattice method
\cite{bard,mat2} in the transverse directions,  A finite lattice
spacing $a$ can be used by choosing the parameters of the
effective theory in a region of renormalization group stability to
respect the required gauge, Poincar\'e, chiral, and continuum
symmetries. The overall normalization gives $f_{\pi} = 101$ MeV
compared with the experimental value of $93$ MeV. Figure
\ref{Fig:DalleyCleo} (a) compares the resulting DLCQ/transverse
lattice pion wavefunction with the best fit to the diffractive
di-jet data (see the next section) after corrections for
hadronization and experimental acceptance \cite{Ashery:1999nq}.
The theoretical curve is somewhat broader than the experimental
result.  However, there are experimental uncertainties from
hadronization and theoretical errors introduced from finite DLCQ
resolution, using a nearly massless pion, ambiguities in setting
the factorization scale $Q^2$, as well as errors in the evolution
of the distribution amplitude from 1 to $10~\mathrm{GeV}^2$.
Instanton models also predict a pion distribution amplitude close
to the asymptotic form \cite{Petrov:1999kg}. In contrast,  recent
lattice results from Del Debbio {\em et al.} \cite{DelDebbio:2000mq}
predict a much narrower shape for the pion distribution amplitude
than the distribution predicted by the transverse lattice. A new
result for the proton distribution amplitude treating nucleons as
chiral solitons has recently been derived by Diakonov and Petrov
\cite{Diakonov:2000pa}. Dyson-Schwinger models \cite{Hecht:2000xa}
of hadronic Bethe-Salpeter wavefunctions can also be used to
predict light-cone wavefunctions and hadron distribution
amplitudes by integrating over the relative $k^-$ momentum.  There
is also the possibility of deriving Bethe-Salpeter wavefunctions
within light-cone gauge quantized QCD \cite{Srivastava:2000gi} in
order to properly match to the light-cone gauge Fock state
decomposition.

\section{Measurements of the Pion Distribution Amplitude by Di-jet
Diffractive Dissociation} The shape of hadron distribution
amplitudes can be measured in the diffractive dissociation of high
energy hadrons into jets on a nucleus. For example, consider the
reaction \cite{Bertsch,MillerFrankfurtStrikman,Frankfurt:1999tq}
$\pi A \rightarrow \mathrm{Jet}_1 + \mathrm{Jet}_2 + A^\prime$ at
high energy where the nucleus $A^\prime$ is left intact in its
ground state.  The transverse momenta of the jets balance so that
$\vec k_{\perp i} + \vec k_{\perp 2} = \vec q_\perp < {R^{-1}}_A$.
The light-cone longitudinal momentum fractions also need to add to
$x_1+x_2 \sim 1$ so that $\Delta p_L < R^{-1}_A$.  The process can
then occur coherently in the nucleus.  Because of color
transparency and the long coherence length,  a valence $q \bar q$
fluctuation of the pion with small impact separation will
penetrate the nucleus with minimal interactions, diffracting into
jet pairs \cite{Bertsch}.  The $x_1=x$, $x_2=1-x$ dependence of
the di-jet distributions will thus reflect the shape of the pion
valence light-cone wavefunction in $x$; similarly, the $\vec
k_{\perp 1}- \vec k_{\perp 2}$ relative transverse momenta of the
jets gives key information on the second derivative of the
underlying shape of the valence pion wavefunction
\cite{MillerFrankfurtStrikman,Frankfurt:1999tq,BHDP}. The
diffractive nuclear amplitude extrapolated to $t = 0$ should be
linear in nuclear number $A$ if color transparency is correct.
The integrated diffractive rate should then scale as $A^2/R^2_A
\sim A^{4/3}$.

The E791 collaboration at Fermilab has recently measured the
diffractive di-jet dissociation of 500 GeV incident pions on
nuclear targets \cite{Ashery:1999nq}.  The results are consistent
with color transparency, and the momentum partition of the jets
conforms closely with the shape of the asymptotic distribution
amplitude, $\phi^\mathrm{asympt}_\pi (x) = \sqrt 3 f_\pi x(1-x)$,
corresponding to the leading anomalous dimension solution
\cite{BrodskyLepage} to the perturbative QCD evolution equation.

The interpretation of the diffractive dijet processes as measures
of the hadron distribution amplitudes has recently been questioned
by Braun {\em et al.} \cite{Braun:2001ih} and by Chernyak
\cite{Chernyak:2001ph} who have calculated the hard scattering
amplitude for such processes at next-to-leading order.  However,
these analyses neglect the integration over the transverse
momentum of the valence quarks and thus miss the logarithmic
ordering which is required for factorization of the distribution
amplitude and color filtering in nuclear targets.

\section {Exclusive Two-Photon Annihilation into Hadron Pairs}

\begin{figure*}[htb]
\begin{center}
\includegraphics[width=.9\textwidth]{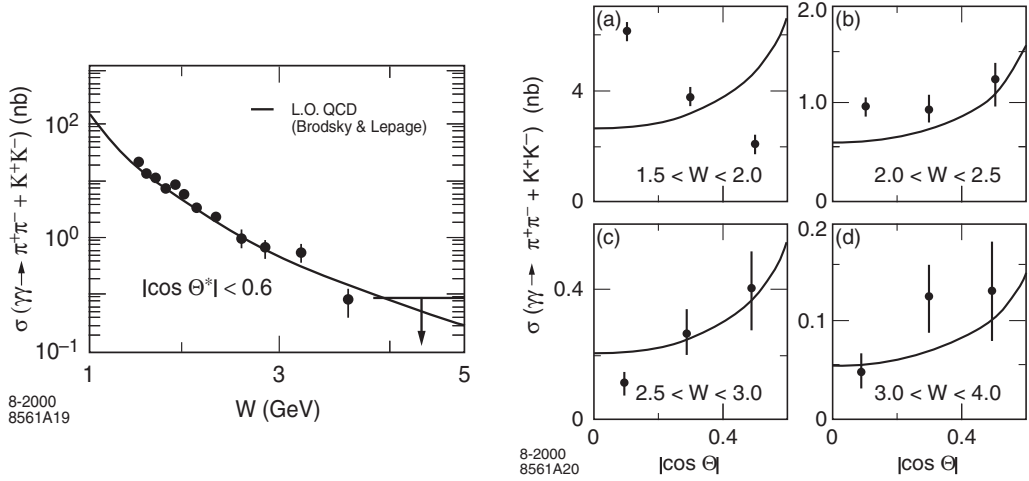}
\end{center}
\caption{Comparison of the sum of $\gamma \gamma \rightarrow \pi^+
\pi^-$ and
$\gamma \gamma \rightarrow K^+ K^-$ meson pair production cross
sections with the scaling and angular distribution of the perturbative QCD
prediction \cite{Brodsky:1981rp}.  The data are from the CLEO
collaboration \cite{Dominick:1994bw}.}
\label{Fig:CLEO}
\end{figure*}

\begin{figure}[htb]
\begin{center}
\includegraphics[width=.48\columnwidth]{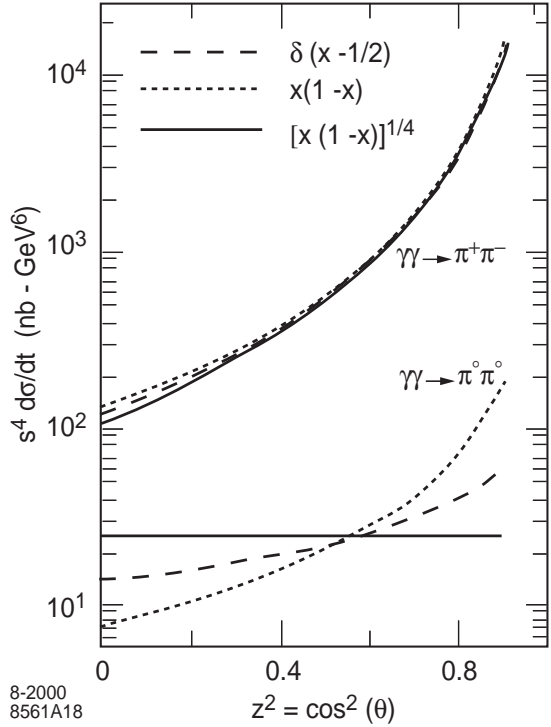}
\end{center}
\caption{Predictions for the angular distribution of the
$\gamma\gamma\rightarrow \pi^+\pi^-$ and
$\gamma \gamma \rightarrow \pi^0 \pi^0$ pair production cross
sections for three different pion distribution
amplitudes \cite{Brodsky:1981rp}.
\label{Fig:piangle}}
\end{figure}

\begin{figure}[htb]
\begin{center}
\includegraphics[width=.48\columnwidth]{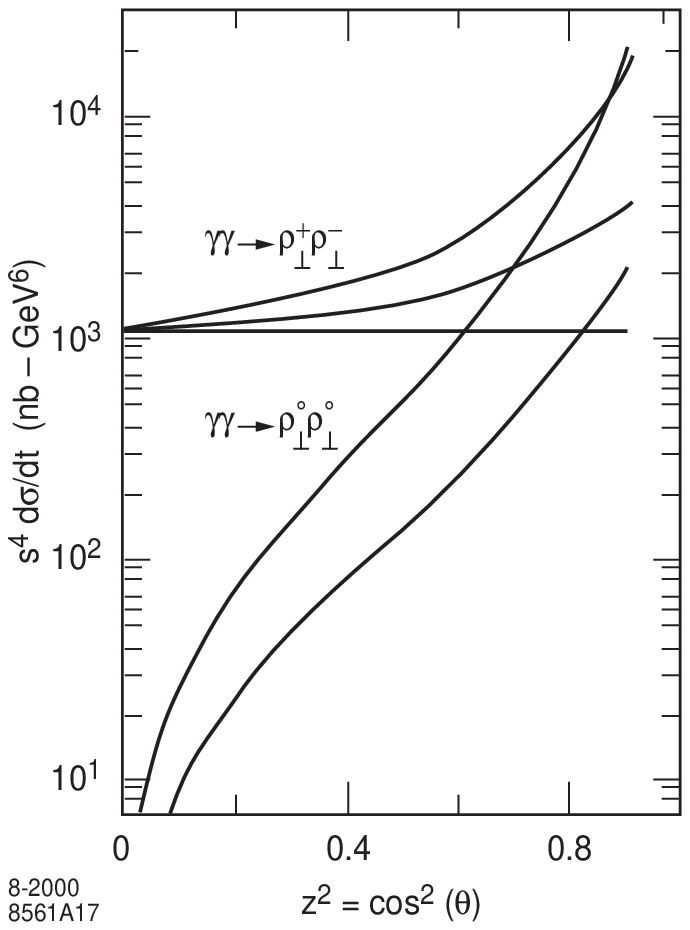}
\end{center}
\caption{Predictions for the angular distribution of the
$\gamma\gamma\rightarrow \rho^+\rho^-$ and
$\gamma \gamma \rightarrow \rho^0 \rho^0$ pair production cross
sections for three different $\rho$ distribution
amplitudes as in Figure \ref{Fig:piangle} \cite{Brodsky:1981rp}.
\label{Fig:rhoangle}}
\end{figure}

Two-photon reactions, $\gamma \gamma \to H \bar H$ at large $s =
(k_1 + k_2)^2$ and fixed $\theta_\mathrm{cm}$, provide a
particularly important laboratory for testing QCD since these
cross-channel ``Compton'' processes are the simplest calculable
large-angle exclusive hadronic scattering reactions. The helicity
structure, and often even the absolute normalization can be
rigorously computed for each two-photon channel
\cite{Brodsky:1981rp}. In the case of meson pairs, dimensional
counting predicts that for large $s$, $s^4 d\sigma/dt(\gamma
\gamma \to M \bar M$ scales at fixed $t/s$ or $\theta_\mathrm{cm}$
up to factors of $\ln s/\Lambda^2$. The angular dependence of the
$\gamma \gamma \to H \bar H$ amplitudes can be used to determine
the shape of the process-independent distribution amplitudes,
$\phi_H(x,Q)$. An important feature of the $\gamma \gamma \to M
\bar M$ amplitude for meson pairs is that the contributions of
Landshoff pitch singularities are power-law suppressed at the Born
level---even before taking into account Sudakov form factor
suppression.  There are also no anomalous contributions from the
$x \to 1$ endpoint integration region. Thus, as in the calculation
of the meson form factors, each fixed-angle helicity amplitude can
be written to leading order in $1/Q$ in the factorized form $[Q^2
= p_T^2 = tu/s$; $\widetilde Q_x = \min(xQ,(l-x)Q)]$:
\begin{eqnarray}
\mathcal{M}_{\gamma \gamma\to M \bar M} &= &\int^1_0\, dx \int^1_0
\, dy \nonumber\\
&&\hspace{-2pc}\phi_{\bar M}(y,\widetilde Q_y) T_H(x,y,s,\theta_\mathrm{cm}
\phi_{M}(x,\widetilde Q_x) ,
\end{eqnarray}
where $T_H$ is the hard-scattering amplitude $\gamma \gamma \to (q
\bar q) (q \bar q)$ for the production of the valence quarks
collinear with each meson, and $\phi_M(x,\widetilde Q)$ is the
amplitude for finding the valence $q$ and $\bar q$ with light-cone
fractions of the meson's momentum, integrated over transverse
momenta $k_\perp < \widetilde Q$. The contribution of non-valence
Fock states are power-law suppressed. Furthermore, the
helicity-selection rules \cite{Brodsky:1981kj} of perturbative QCD
predict that vector mesons are produced with opposite helicities
to leading order in $1/Q$ and all orders in $\alpha_s$. The
dependence in $x$ and $y$ of several terms in $T_{\lambda,
\lambda'}$ is quite similar to that appearing in the meson's
electromagnetic form factor. Thus much of the dependence on
$\phi_M(x,Q)$ can be eliminated by expressing it in terms of the
meson form factor. In fact, the ratio of the $\gamma \gamma \to
\pi^+ \pi^-$ and $e^+ e^- \to \mu^+ \mu^-$ amplitudes at large $s$
and fixed $\theta_{CM}$ is nearly insensitive to the running
coupling and the shape of the pion distribution amplitude:
\begin{equation}{{d\sigma \over dt }(\gamma \gamma \to \pi^+ \pi^-)
\over {d\sigma \over dt }(\gamma \gamma \to \mu^+ \mu^-)} \sim {4
\vert F_\pi(s) \vert^2 \over 1 - \cos^2 \theta_\mathrm{cm} }.
\end{equation}
The comparison of the PQCD prediction for the sum
of $\pi^+ \pi^-$ plus $K^+ K^-$ channels with recent CLEO data
\cite{Dominick:1994bw} is shown in Figure \ref{Fig:CLEO}. The CLEO data for
charged pion and kaon pairs show a clear transition to the scaling
and angular distribution predicted by PQCD \cite{Brodsky:1981rp}
for $W = \sqrt(s_{\gamma \gamma} > 2$ GeV.  It is clearly
important to measure the magnitude and angular dependence of the
two-photon production of neutral pions and $\rho^+ \rho^-$ in view
of the strong sensitivity of these channels to the shape of meson
distribution amplitudes (see Figures \ref{Fig:piangle} and
\ref{Fig:rhoangle}). QCD also predicts that the production cross
section for charged $\rho$-pairs (with any helicity) is much
larger than for that of neutral $\rho$ pairs, particularly at
large $\theta_\mathrm{cm}$ angles. Similar predictions are possible
for other helicity-zero mesons.

\begin{figure*}[htbp]
\begin{center}
\includegraphics[width=5.5in]{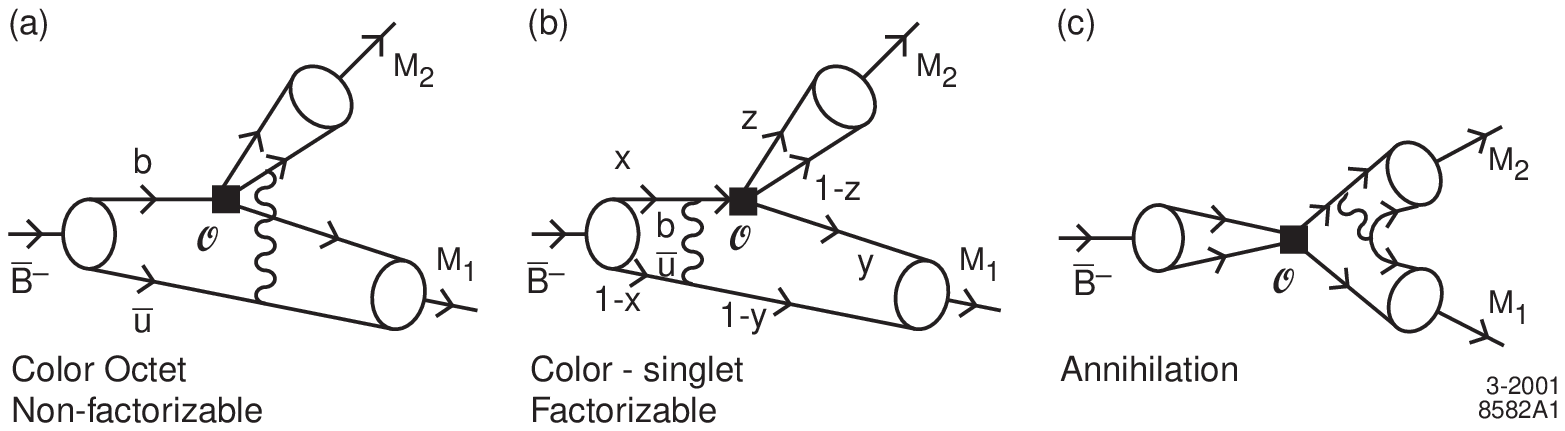}
\end{center}
\caption{Three representative contributions to exclusive $B$ decays to
meson pairs in PQCD.  The operators $\cal O$ represent the QCD-improved
effective weak interaction.
\label{fig:B}}
\end{figure*}

The analysis of exclusive $B$ decays has much in common with
the analysis of exclusive two-photon reactions \cite{Brodsky:2001jw}.
For example, consider the
three representative contributions to the decay of a $B$ meson to meson
pairs illustrated in Figure  \ref{fig:B}.  In Figure \ref{fig:B}(a) the weak
interaction effective operator $\mathcal{O}$ produces a $ q \bar q$ in a
color octet state.  A gluon with virtuality
$Q^2 = \mathcal{O} (M_B^2)$ is needed to equilibrate the large
momentum fraction carried by the $b$ quark in the $\bar B$ wavefunction.
The amplitude then factors into a hard QCD/electroweak subprocess
amplitude for quarks which are collinear with their respective
hadrons:
$T_H([b(x)
\bar u(1-x)]
\to [q(y) \bar u(1-y)]_1 [q(z) \bar q(1-z)]_2)$
convoluted with the distribution amplitudes $\phi(x,Q)$
\cite{BrodskyLepage} of the
incident and final hadrons:
\begin{eqnarray*}
\mathcal{M}_\mathrm{octet}(B \to M_1 M_2) &= &\int^1_0\, dz \int^1_0\, dy \int^1_0\,
dx\\
&&\hspace{-3pc}\phi_B(x,Q) T_H(x,y,z) \phi_{M_1}(y,Q) \phi_{M_2}(z,Q).
\end{eqnarray*}
Here $x = k^{+}/p^{+}_H = (k^0+ k^z) /(p^0_H + p^z_H)$ are the
light-cone momentum fractions carried by the valence quarks.

There are a several features of QCD which are required to
ensure the consistency of the PQCD approach: (a) the
effective QCD coupling
$\alpha_s(Q^2)$ needs to be under control
at the relevant scales of $B$ decay;
(b) the distribution amplitudes of the hadrons need
to satisfy convergence properties at the endpoints; and (c) one requires the
coherent cancelation of the couplings of soft gluons to color-singlet states.
This property, color transparency \cite{Brodsky:1988xz}, is a fundamental
coherence property of gauge theory and leads to diminished final-state
interactions and corrections to the PQCD factorizable contributions.  The
problem of setting the renormalization scale of the coupling for exclusive
amplitudes is discussed in  \cite{Brodsky:1998dh}.

Baryon pair production in two-photon annihilation is also an
important testing ground for QCD.  The calculation of $T_H$ for
Compton scattering requires the evaluation of 368
helicity-conserving tree diagrams which contribute to $\gamma
(qqq) \to \gamma^\prime (qqq)^\prime$ at the Born level and a
careful integration over singular intermediate energy denominators
\cite{Farrar:1990qj,Kronfeld:1991kp,Guichon:1998xv}. Brooks and
Dixon \cite{Brooks:2000nb} have recently completed a recalculation
of the proton Compton process at leading order in PQCD, extending
and correcting earlier work.  It is useful to consider the ratio $
d\sigma/dt(\gamma \gamma \to \bar p p)/ d\sigma/dt(e^+ e^- \to
\bar p p)$ since the power-law fall-off, the normalization of the
valence wavefunctions, and much of the uncertainty from the scale
of the QCD coupling cancel. The scaling and angular dependence of
this ratio is sensitive to the shape of the proton distribution
amplitudes. The perturbative QCD predictions for the phase of the
Compton amplitude phase can be tested in virtual Compton
scattering by interference with Bethe-Heitler processes
\cite{Brodsky:1972vv}.

It is also interesting to measure baryon and isobar pair
production in two photon reactions near threshold.  Ratios such as
$\sigma(\gamma \gamma \to \Delta^{++} \Delta^{--})/\sigma(\gamma
\gamma \to \Delta^{+} \Delta^{-})$ can be as large as $16:1$ in
the quark model since the three-quark wavefunction of the $\Delta$
is expected to be symmetric \cite{Karliner}. Such large ratios
would not be expected in soliton models \cite{Sommermann:1992yh}
in which intermediate multi-pion channels play a major role.

Recently Pobylitsa  {\em et al.} \cite{Pobylitsa:2001cz} have shown how
the predictions of perturbative QCD can be extended to processes
such as $\gamma \gamma \to p \bar p \pi$ where the pion is
produced at low velocities relative to that of the $p$ or $\bar p$
by utilizing soft pion theorems in analogy to soft photon theorems
in QED.  The distribution amplitude of the $p \pi$ composite is
obtained from the proton distribution amplitude from a chiral
rotation.  A test of this procedure in inelastic electron
scattering at large momentum transfer $e p \to p \pi$ and small
invariant $p^\prime \pi$ mass has been remarkably successful. Many
tests of the soft meson procedure are possible in multiparticle
$e^+ e^-$ and $\gamma \gamma$ final states.

\section{Conclusions}

The leading-twist QCD predictions for exclusive two-photon processes such
as the photon-to-pion transition form factor and $\gamma \gamma \to $
hadron pairs are based on rigorous factorization theorems.  The recent
data from the CLEO collaboration on $F_{\gamma \pi}(Q^2)$ and the sum of
$\gamma \gamma \to \pi^+ \pi^-$ and $\gamma \gamma \to K^+ K^-$
channels are in excellent agreement with the QCD predictions.  It is
particularly compelling to see a transition in angular dependence between
the low energy chiral and PQCD regimes.  The success of leading-twist
perturbative QCD scaling
for exclusive processes at presently experimentally accessible momentum
transfer can be understood if the effective coupling
$\alpha_V(Q^*)$ is approximately constant at the relatively
small scales $Q^*$ relevant to the hard scattering
amplitudes \cite{Brodsky:1998dh}.  The evolution of the quark distribution
amplitudes in the low-$Q^*$ domain also needs to be minimal.  Sudakov
suppression of the endpoint contributions is also strengthened if the
coupling is frozen because of the exponentiation of a double logarithmic
series.

One of the formidable challenges in QCD is the calculation of
non-perturbative wavefunctions of hadrons from first principles.  The
recent calculation of the pion distribution amplitude by
Dalley \cite{Dalley:2000dh}
and by Burkardt and Seal \cite{Burkardt:2001mf}
using light-cone and transverse lattice methods is
particularly encouraging.  The predicted form of
$\phi_\pi(x,Q)$ is somewhat broader than but not inconsistent with the
asymptotic form favored by the measured normalization of $Q^2 F_{\gamma
\pi^0}(Q^2)$ and the pion wavefunction inferred from diffractive di-jet
production.

Clearly much more experimental input on hadron wavefunctions is
needed, particularly from measurements of two-photon exclusive
reactions into meson and baryon pairs at the high luminosity $B$
factories.  For example, as shown in Figure \ref{Fig:piangle}, the
ratio
\begin{displaymath}
{{d\sigma \over dt }(\gamma \gamma \to \pi^0 \pi^0) /
{d\sigma \over dt}(\gamma \gamma \to \pi^+ \pi^-)}
\end{displaymath}
is particularly sensitive to the shape of pion distribution
amplitude. At fixed pair mass, and high photon virtuality, one can
study the distribution amplitude of multi-hadron states
\cite{Diehl:2000uv}. Two-photon annihilation will provide much
information on fundamental QCD processes such as deeply virtual
Compton scattering and large angle Compton scattering in the
crossed channel.  I have also emphasized the interrelation between
the wavefunctions measured in two-photon collisions and the
wavefunctions needed to study exclusive $B$ and $D$ decays.

Much of the most interesting
two-photon annihilation physics is accessible at low energy,  high
luminosity $e^+ e^-$ colliders,
including measurements of channels important in the light-by-light
contribution to the muon
$g$--2 and the study of the transition between threshold production
controlled by low-energy effective chiral theories and the domain where
leading-twist perturbative QCD becomes applicable.

The threshold regime
of hadron production in photon-photon and $e^+ e^-$ annihilation, where
hadrons are formed at small relative velocity, is particularly interesting
as a test of low energy theorems, soliton models, and new types of
resonance production.  Such studies will be particularly valuable in
double-tagged reactions where polarization correlations, as well as the
photon virtuality dependence, can be studied.

\section*{Acknowledgments}

Work supported by the Department of Energy
under contract number DE-AC03-76SF00515.

\end{document}